\documentclass{iau}
\usepackage{graphicx} 

\title[IAUS291.~~Maximum mass of neutron stars with quark cores] 
{Does a hadron-quark phase transition in dense matter preclude the existence of massive neutron stars ?} 

\author[N. Chamel et al.]  
{N. Chamel$^1$,
A.F. Fantina$^1$, J.M. Pearson$^2$ \and S. Goriely$^1$}

\affiliation{$^1$Institut d'Astronomie et d'Astrophysique, CP-226, Universit\'e Libre de Bruxelles, 1050 Brussels, Belgium \\[\affilskip]
$^2$D\'ept. de Physique, Universit\'e de Montr\'eal, Montr\'eal (Qu\'ebec), H3C 3J7 Canada}

\pubyear{2012}
\volume{291}  
\jname{\mbox{Neutron Stars and Pulsars: Challenges and Opportunities after 80 years}}
\editors{J. van Leeuwen, ed.} 
\begin{document}

\maketitle

\begin{abstract}
We study the impact of a hadron-quark phase transition on the maximum neutron-star mass. 
The hadronic part of the equation of state relies on the most up-to-date Skyrme nuclear energy 
density functionals, fitted to essentially all experimental nuclear mass data and constrained to reproduce 
the properties of infinite nuclear matter as obtained from microscopic calculations using realistic forces. 
We show that the softening of the dense matter equation of state due to the phase transition is not necessarily 
incompatible with the existence of massive neutron stars like PSR J1614$-$2230.
\end{abstract}


\firstsection 
\section{Introduction}

Neutron stars (NSs) result from the gravitational collapse of massive stars with $M \gtrsim 8 M_\odot$ at the end point 
of their evolution. They are among the most compact objects in the universe, with a central density which can reach several 
times the nuclear saturation density. At least, three different regions can be identified in the interior of a NS: 
(i) the ``outer crust'', at densities above $\sim 10^4$ g~cm$^{-3}$, composed of fully ionized atoms, arranged in a Coulomb 
lattice of nuclei, neutralized by a degenerate electron gas, (ii) the ``inner crust'', at densities above $\sim 4 \times 10^{11}$~g~cm$^{-3}$, 
composed of neutron-proton clusters and unbound neutrons, neutralized by a degenerate electron gas, and (iii) the core, at densities 
above $\sim 10^{14}$ g~cm$^{-3}$. The precise measurement of the mass of pulsar PSR J1614$-$2230 by \cite{demorest2010} has revived 
the question of the composition of the core. Just below the crust, the matter consists of a mixture of neutrons, protons, electrons 
and possibly muons. The composition of the central region of a NS is still a matter of debate (see e.g. \cite{haensel2007}). 

In the present work, we study the impact of a hadron-quark phase transition in dense matter on the maximum mass of cold isolated NSs 
(see \cite{chamel2013} for a general discussion of the maximum mass of hybrid stars).

\section{Hadronic equation of state}

The global structure of a NS is determined by the equation of state (EoS), i.e. the relation between the matter pressure $P$ and the 
mass-energy density $\rho$. Before considering the possibility of a phase transition from hadronic to quark matter in the core of NSs, we 
will begin with the hadronic EoSs. A good starting point is the family of three EoSs, BSk19, BSk20 and BSk21, which have been developed to 
provide a unified treatment of all regions of a NS (see \cite{pearson2011, pearson2012}). These EoSs are based on nuclear energy-density 
functionals derived from generalized Skyrme forces (in that they contain additional momentum- and density-dependent terms), which fit 
essentially all measured masses of atomic nuclei with an rms deviation of 0.58 MeV for all three models. Moreover, these functionals were 
constrained to reproduce three different neutron matter EoSs, as obtained from microscopic calculations (see \cite{goriely2010}). All three 
EoSs assume that the core of a NS is made of nucleons and leptons. The BSk19 EoS was found to be too soft to support NSs as massive as PSR 
J1614$-$2230 (\cite{chamel2011}) and therefore, it will not be considered here.

\section{Hadron-quark phase transition}

Given the uncertainties in the composition of dense matter in NSs, we will simply suppose that above the average baryon density $n_{\rm N}$, 
matter undergoes a first-order phase transition to deconfined quark matter subject to the following restrictions: 
(i) for the transition to occur the energy density of the quark phase must be lower than that of the hadronic phase, 
(ii) according to perturbative quantum chromodynamics (QCD) calculations (e.g. \cite{kurkela2010}), the speed of sound in quark matter cannot exceed $c/\sqrt{3}$ 
where $c$ is the speed of light. At densities below $n_{\rm N}$, matter is purely hadronic while a pure quark phase is found at densities 
above some density $n_{\rm X}$. In the intermediate region ($n_{\rm N}<n<n_{\rm X}$) where the two phases can coexist, the pressure and the 
chemical potential of the two phases are equal: $P_{\rm quark}(n) = P_{\rm hadron}(n_N)$ and $\mu_{\rm quark}(n) = \mu_{\rm hadron}(n_N)$.
The EoS of the quark phase at $n>n_{\rm X}$ is given by:
\begin{equation}
P_{\rm quark}(n) = \frac{1}{3} (\mathcal{E}_{\rm quark}(n)-\mathcal{E}_{\rm quark}(n_{\rm X})) + P_{\rm hadron}(n_{\rm N}) \, .
\label{eq:quark}
\end{equation}
We set the density $n_{\rm N}$ to lie above the highest density found in nuclei as predicted by Hartree-Fock-Bogoliubov calculations, 
namely $n_{\rm N}=0.2$~fm$^{-3}$ (\cite{bruslib}). The density $n_{\rm X}$ is adjusted to optimize
the maximum mass under the conditions mentioned above. Eq.~(\ref{eq:quark}) turns out to be very similar to that obtained within the simple 
MIT bag model, which has been widely applied to describe quark matter in compact stars (see e.g. \cite{haensel2007}).
The effective bag constant $B$ associated with the BSk21 hadronic EoS is $56.7$~MeV~fm$^{-3}$.

\section{Maximum mass} 

\begin{figure}[t]
\centering
\includegraphics[scale=0.3]{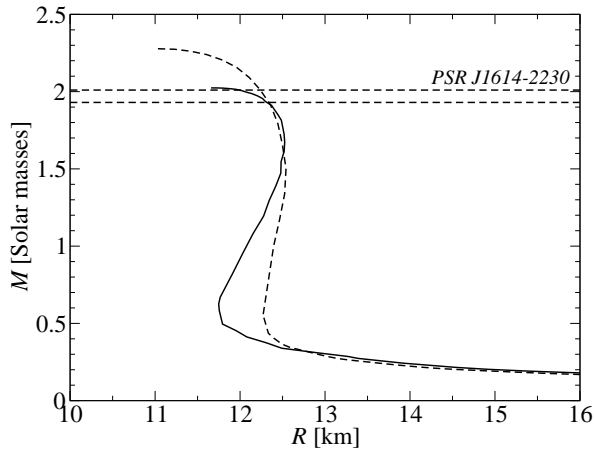}
\caption{Gravitational mass versus circumferential radius with (solid line) and without (dashed line) a quark-matter core. See the text for 
detail.}
\label{fig01}
\end{figure}

Considering the stiffest hadronic EoS (BSk21), we have solved the Tolman-Oppenheimer-Volkoff equations (\cite{tolman1939, oppenheimervolkoff1939}) 
in order to determine the global structure of a nonrotating NS. The effect of rotation on the maximum mass was found to be
very small for stars with spin-periods comparable to that of PSR J1614$-$2230 (\cite{chamel2011}); we therefore neglect it. The gravitational 
mass versus circumferential radius relation is shown in Fig.~\ref{fig01}. We have considered two cases: a purely hadronic NS 
described by our BSk21 EoS (dashed line) and a hybrid star with a quark core (solid line). The corresponding maximum masses are 2.28~$M_\odot$ 
and 2.02~$M_\odot$ respectively. In both cases, the existence of two-solar mass NSs is therefore allowed. 

\section{Conclusions}

The presence of a deconfined quark-matter phase in NS cores leads to a maximum mass of about $2M_\odot$, which is still compatible with 
the mass measurement of PSR J1614$-$2230 by \cite{demorest2010}, but which could be challenged by observations of significantly more 
massive NSs (see \cite{cla02,fre08,kbk11}) unless the sound speed in quark matter is significantly larger than that predicted by 
perturbative QCD calculations (\cite{kurkela2010}).

\acknowledgments FNRS (Belgium), NSERC (Canada) and CompStar, a Research Networking Programme of the European 
Science Foundation are gratefully acknowledged.


\begin{thebibliography}{}
\bibitem[BRUSLIB]{bruslib} BRUSLIB http://www.astro.ulb.ac.be/bruslib
\bibitem[Chamel et al. 2011]{chamel2011} Chamel, N., Fantina, A.~F., Pearson, J.~M., \& Goriely, S.\ 2011, \textit{Phys. Rev. C}, 84, 062802 
\bibitem[Chamel et al. 2013]{chamel2013} Chamel, N., Fantina, A.~F., Pearson, J.~M., \& Goriely, S.\ 2013, \textit{Astron. Astrophys.}, 553, A22
\bibitem[Clark et al. 2002]{cla02} Clark, J.~S., Goodwin, S.~P., Crowther,  P.~A., Kaper, L,  Fairbairn, M.,  Langer, N., Brocksopp, C. \ 2002, \textit{Astron. Astrophys.}, 392, 909
\bibitem[Demorest et al. (2010)]{demorest2010} Demorest, P.~B., Pennucci, T., Ransom, S.~M., Roberts, M.~S.~E., \& Hessels, J.~W.~T.\ 2010, \textit{Nature}, 467, 1081 
\bibitem[Freire et al. 2008]{fre08} Freire, P.~C.~C., Ransom, S.~M., B\'egin, S., Stairs, I.~H., Hessels, J.~W.~T., Frey, L.~H., Camilo, F. \ 2008, \textit{Astrophys. J.}, 675, 670
\bibitem[Goriely et al. 2010]{goriely2010} Goriely, S., Chamel, N., \& Pearson, J.~M.\ 2010, \textit{Phys. Rev. C}, 82, 035804 
\bibitem[Haensel et al. 2007]{haensel2007} Haensel, P., Potekhin, A.~Y., \& Yakovlev, D.~G.\ 2007, \textit{Astrophys. Space Sc. L.}, 326 
\bibitem[Kurkela et al. 2010]{kurkela2010} Kurkela, A., Romatschke, P., \& Vuorinen, A.\ 2010, \textit{Phys. Rev. D}, 81, 105021 
\bibitem[Oppenheimer \& Volkoff 1939]{oppenheimervolkoff1939} Oppenheimer, J.~R., \& Volkoff, G.~M.\ 1939, \textit{Phys. Rev.}, 55, 374 
\bibitem[Pearson et al. 2011]{pearson2011} Pearson, J.~M., Goriely, S., \& Chamel, N.\ 2011, \textit{Phys. Rev. C}, 83, 065810 
\bibitem[Pearson et al. 2012]{pearson2012} Pearson, J.~M., Chamel, N., Goriely, S., \& Ducoin, C.\ 2012, \textit{Phys. Rev. C}, 85, 065803
\bibitem[Tolman 1939]{tolman1939} Tolman, R.~C.\ 1939, \textit{Phys. Rev.}, 55, 364 
\bibitem[van Kerkwijk et al. 2011]{kbk11} van Kerkwijk, M.~H., Breton, R.~P.,  Kulkarni, S.~R.\ 2011, \textit{Astrophys. J.} 728, 95
\end{thebibliography}
\end{document}